\documentclass[aps,prb,10pt,twocolumn,superscriptaddress]{revtex4}
\usepackage{amsmath,amssymb,bm}
\usepackage{graphicx}
\usepackage{epstopdf}
\usepackage{latexsym}
\usepackage{adjustbox}
\usepackage{float}
\usepackage{dsfont}	
\usepackage{makecell} 
\usepackage{color}
\usepackage{natbib}
\usepackage{braket}
\usepackage{hyperref}
\usepackage{comment}
\usepackage{array} 
\hypersetup{
  colorlinks,
  citecolor=magenta,
  linkcolor=blue,
  urlcolor=blue}

\usepackage{tikz}
\newcommand{\thickbond}{%
\tikz[baseline]{\draw[line width=0.5mm] (0,-0.3mm) -- (0,0.15cm); \fill[violet] (0,-0.3mm) circle (0.4mm); \fill[violet] (0,0.15cm) circle (0.4mm)}%
} 

\newcommand{\thinbond}{%
\tikz[baseline]{\draw[line width=0.2mm] (0,-0.3mm) -- (0,0.15cm); \fill[violet] (0,-0.3mm) circle (0.4mm); \fill[violet] (0,0.15cm) circle (0.4mm)}%
}

\newcommand{\myarrow}{%
\tikz[baseline]{ \draw [->, >=latex, very thick] (0,0) -- (1cm,0);}
}

\begin{document}

\title{Reduction of the sign problem near $T=0$ in quantum Monte Carlo simulations}

\author{Jonathan D'Emidio}
\affiliation{Institute of Physics, \'{E}cole Polytechnique F\'{e}d\'{e}rale de Lausanne (EPFL), CH-1015 Lausanne, Switzerland}
\author{Stefan Wessel}
\affiliation{Institut f\"ur Theoretische Festk\"orperphysik, JARA-FIT and JARA-HPC, RWTH Aachen University, 52056 Aachen, Germany}
\author{Fr\'{e}d\'{e}ric Mila}
\affiliation{Institute of Physics, \'{E}cole Polytechnique F\'{e}d\'{e}rale de Lausanne (EPFL), CH-1015 Lausanne, Switzerland}

\begin{abstract}
Building on a recent investigation of the Shastry-Sutherland model [S. Wessel \textit{et al}., Phys. Rev. B \textbf{98}, 174432 (2018)], we develop a general strategy to eliminate the Monte Carlo sign problem near the zero temperature limit in frustrated quantum spin models.  If the Hamiltonian of interest and the sign-problem-free Hamiltonian---obtained by making all off-diagonal elements negative in a given basis---have the same ground state and this state is a member of the computational basis, then the average sign returns to one as the temperature goes to zero.  We illustrate this technique by studying the triangular and kagome lattice Heisenberg antiferrromagnet in a magnetic field above saturation, as well as the Heisenberg antiferromagnet on a modified Husimi cactus in the dimer basis.  We also provide detailed appendices on using linear programming techniques to automatically generate efficient directed loop updates in quantum Monte Carlo simulations.
\end{abstract}
\date{\today}
\maketitle

\section{Introduction}

The sign problem represents the biggest obstacle in applying Monte Carlo techniques to resolve quantum many-body problems.  When present, it renders Monte Carlo as a method that trades the exponential scaling of the Hilbert space for an exponential scaling of the simulation runtime.  Even though a general solution to the sign problem is unlikely [\onlinecite{Troyer2005:CompComplx}], there exist many examples where the sign problem has been overcome in specific models [\onlinecite{Chandrasekharan1999:Meron, Henelius2000:SignProbFrustSpin, Wu2003:ExactSO5, Nyfeler2008:NestedCluster, Okunishi2014:SymProtSign, Li2016:Majorana, Hann2017:SignProbImp, Nakamura1998:VanishSign, Alet2016:SignProbFree, Honecker2016:ThermodynBound}].

One strategy that has been highly successful in eliminating the sign problem from certain classes of frustrated magnets has been the use of the dimer basis [\onlinecite{Nakamura1998:VanishSign, Alet2016:SignProbFree, Honecker2016:ThermodynBound, Wessel2017:EfficientQMC, Ng2017:FrustSpinDimer, Wessel2018:ThermoShastry, Stapmanns2018:TCPBilayer}].  This has resulted in the ability to efficiently simulate frustrated 1D ladder systems [\onlinecite{Nakamura1998:VanishSign, Alet2016:SignProbFree, Honecker2016:ThermodynBound, Wessel2017:EfficientQMC}] and 2D bilayer systems [\onlinecite{Alet2016:SignProbFree, Ng2017:FrustSpinDimer, Stapmanns2018:TCPBilayer}].  This technique was also applied to frustrated systems such as the Shastry Sutherland model [\onlinecite{Shastry1981:SSM}] that still suffer from a sign problem even in the dimer basis [\onlinecite{Wessel2018:ThermoShastry}].  However, here it was noted that the sign problem is remarkably mild and even disappears at low temperatures in the dimer singlet phase.  A similar reduction of the sign problem was previously observed in frustrated ladder systems in the dimer basis [\onlinecite{Wessel2017:EfficientQMC}].  This motivates the investigation of whether or not such a reduction in the sign problem could generically be used to efficiently study the physics of highly frustrated quantum systems.

To address this question, our goal in this work is to explore the easing of the sign problem at low temperatures in models of frustrated magnets.  We will specify the conditions that are necessary for this situation to arise (as was pointed out in [\onlinecite{Wessel2018:ThermoShastry}]), and explore two models that demonstrate this behavior.  We use two-dimensional frustrated Heisenberg models in a large magnetic field in the $S^z$ basis as the simplest illustration of the effect.  For this case we also provide a comparison of our statistically exact thermodynamic measurements with a mean field treatment of the problem.  We then move to the dimer basis and study the Heisenberg antiferromagnet on a modified Husimi cactus [\onlinecite{Zeng1990:KagomeNet, Elser1993:KagomeHyperbolic, Chandra1994:Husimi}], which locally mimics the kagome lattice.  In Appendix \ref{appendix:DimerHam} we provide details on the Husimi Heisenberg Hamiltonian in the dimer basis. Finally in Appendix \ref{appendix:LinearProg} we give explicit instructions on the linear programming technique suggested in [\onlinecite{Alet2005:GenDirLoop}], which greatly facilitates the implementation of directed loop updates [\onlinecite{Syljuasen2002:DirectedLoops}] in the dimer basis.  This technique can be used to fully automate the simulation of arbitrarily complicated models, requiring only the local Hamiltonian matrix as an input.

\section{\label{sec:Gen} General conditions}

Here we outline the general conditions that will result in an easing of the average sign at low temperatures, as was first pointed out in [\onlinecite{Wessel2018:ThermoShastry}].  We first require that the ground state of the Hamiltonian of interest is a member of the computational basis being used.  Secondly, we require that this state is also the ground state of the sign-problem-free Hamiltonian obtained by making all off-diagonal matrix elements negative.  The sign-problem-free Hamiltonian is thus defined with respect to the basis being used.  With these two conditions met, the ground state energy of the originally signed Hamiltonian ($H^{\texttt{-}}$) and the sign-problem-free Hamiltonian ($H^{\texttt{+}}$) will be the same, leading to an average sign that returns to one at low temperatures.

The average sign can be computed as a ratio of partition functions that, using the fact that the ground state energies are the same, can be expressed as

\begin{equation}
\label{eq:avgsign}
\langle \mathrm{sign} \rangle = \frac{\text{Tr}(e^{-\beta H^{\texttt{-}}})}{\text{Tr}(e^{-\beta H^{\texttt{+}}})} = \frac{1+\sum_{i}e^{-\beta \Delta^{\texttt{-}}_i}}{1+\sum_{i}e^{-\beta \Delta^{\texttt{+}}_i}},
\end{equation}
where $\beta=1/T$ is the inverse temperature and $\Delta^{\texttt{+}}_i$, $\Delta^{\texttt{-}}_i$ are the energy gaps for all excited states (relative to the ground state energy) in the spectrum of $H^{\texttt{+}}$, $H^{\texttt{-}}$, respectively.  From here we can clearly see that $\langle \text{sign} \rangle \to 1$ as $T \to 0$.  Also, increasing the size of the gaps will equally cause an easing of the sign.  We will now demonstrate the utility of these principles to study the thermodynamics of frustrated Heisenberg models.

\section{\label{sec:Bfield} Frustrated Heisenberg antiferromagnets in external field}

\begin{figure}[!t]
\centerline{\includegraphics[angle=0,width=1.0\columnwidth]{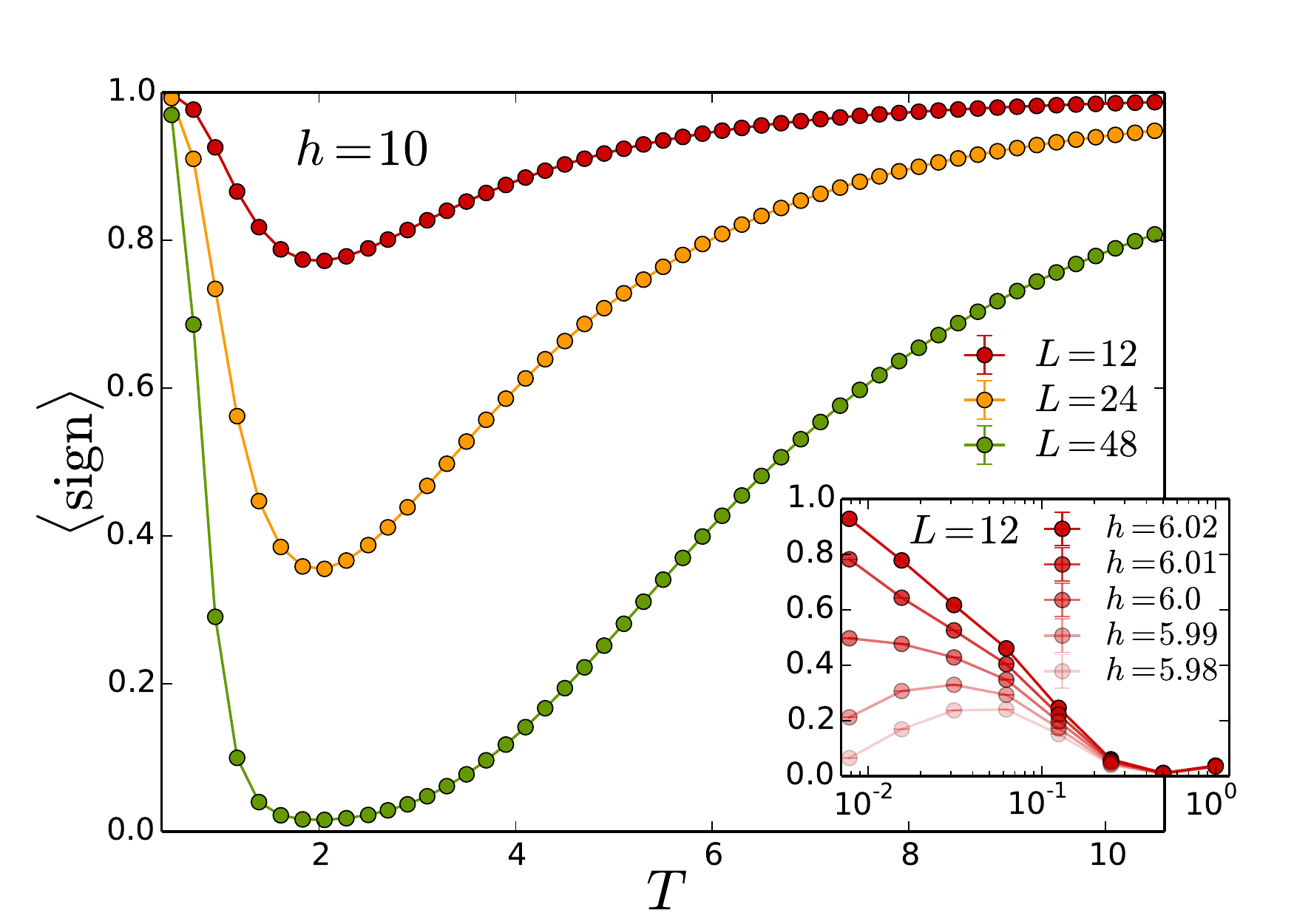}}
\caption{The average sign as a function of temperature for the triangular lattice Heisenberg antiferromagnet in a large external field ($h=10$).  At fields above the saturation threshold for $H^{\texttt{+}}$ ($h^{\texttt{+}}_s=6$), which is above the threshold for $H^{\texttt{-}}$ ($h^{\texttt{-}}_s=4.5$), the sign returns to one at low temperatures.}
\label{fig:TriSign}
\end{figure}

As the most simple demonstration of this effect, we first consider frustrated Heisenberg antiferromagnets on the triangular and kagome lattices in a strong magnetic field.  Taking the computational basis to be the standard $S^z$ spin values, we can write the Hamiltonians as

\begin{equation}
\label{eq:Hpm}
H^{\mp} = J \sum_{\langle i j \rangle} \left(S^z_i S^z_j \pm \frac{1}{2}(S^+_i S^-_j + S^-_i S^+_j ) \right) - h\sum_i S^z_i.
\end{equation}

It is clear that $H^{\texttt{+}}$ does not suffer from a sign problem since the off diagonal matrix elements are all negative.  And importantly, for large enough values of the external field, $H^{\texttt{+}}$ and $H^{\texttt{-}}$ will both have the fully polarized state as the ground state with the same energy.  Thus in this limit we expect an easing of the average sign as ensured by Eq. (\ref{eq:avgsign}).

  We have employed the stochastic series expansion algorithm [\onlinecite{Sandvik2010:CompStud}] using directed loops [\onlinecite{Syljuasen2002:DirectedLoops}] to compute the finite temperature properties of the Hamiltonian in Eq. (\ref{eq:Hpm}).  In Fig. \ref{fig:TriSign} we show the temperature dependence of the average sign on $L \times L$ triangular lattices in a large external field, $h=10$ (setting $J=1$).  To illustrate the point, we have chosen the field to be well above the saturation thresholds $h^{\texttt{-}}_s=4.5$ and $h^{\texttt{+}}_s=6$ for $H^{\texttt{-}}$ and $H^{\texttt{+}}$, respectively.  One clearly sees an intermediate temperature scale where the sign goes to zero upon increasing the system size, however at sufficiently low temperatures the sign goes back to one as expected.

\begin{figure}[t!]
\centerline{\includegraphics[angle=0,width=1.0\columnwidth]{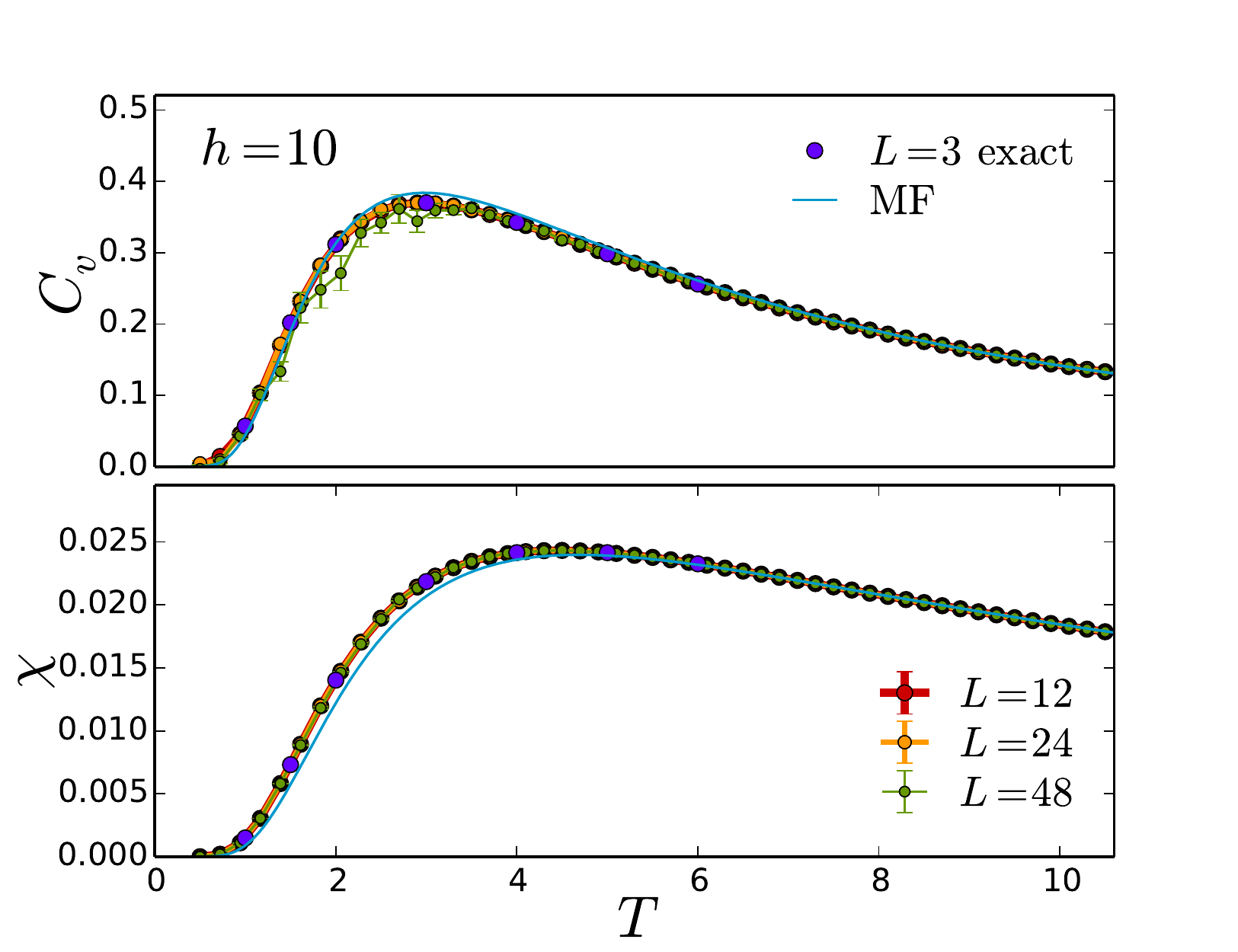}}
\caption{The specific heat and magnetic susceptibility measured in the same simulations as Fig. \ref{fig:TriSign}.  We note that the finite size effects in this large field are essentially negligible, and even the $L=3$ system from exact diagonalization agrees with the thermodynamic limit.  The numerical data is also very close to the mean field prediction (MF).}
\label{fig:TriChi}
\end{figure}

In Fig. \ref{fig:TriChi} we show the specific heat and magnetic susceptibility measured in the same runs as in Fig. \ref{fig:TriSign}.  Most notably, we see that the data shows extremely small finite size effects, such that even the $L=3$ data obtained by exact diagonalization is essentially converged to the thermodynamic limit.  In fact a simple mean field treatment of the problem also gives very good agreement at such high fields, as also depicted in Fig. \ref{fig:TriChi}.  In the following section we outline how this mean field prediction was obtained and compare these results to our numerical data at smaller fields, studying as well the finite size dependence.

\section{\label{sec:MF}Mean field treatment}

We now perform a standard mean field decoupling of the Heisenberg antiferromagnet in an external field, leading to a mean field Hamiltonian of independent spins in an effective field:

\begin{equation}
\label{eq:HMF}
H_{\text{MF}} = -\frac{J N_s N_c m^2}{2} - h_m \sum_i S^z_i.
\end{equation}
Here $N_s$ is the number of sites, $N_c$ is the coordination number of the lattice, $m\equiv \langle S^z_i \rangle$ is the magnetization per site, and $h_m=h - J N_c m$ is the effective field.  The self consistency condition is given by

\begin{equation}
\label{eq:MFsc}
m = \tfrac{1}{2} \tanh(\tfrac{\beta h_m}{2}).
\end{equation}
Given this, one can derive the form of $C_v$ and $\chi$:

\begin{equation}
\label{eq:Cv}
C_v = \frac{\beta^2 h^2_m}{4\cosh^{2}(\tfrac{\beta h_m}{2})+\beta J N_c},
\end{equation}

\begin{equation}
\label{eq:Chi}
\chi = \frac{\beta}{4\cosh^{2}(\tfrac{\beta h_m}{2})+\beta J N_c}.
\end{equation}

\begin{figure}[h!]
\centerline{\includegraphics[angle=0,width=1.0\columnwidth]{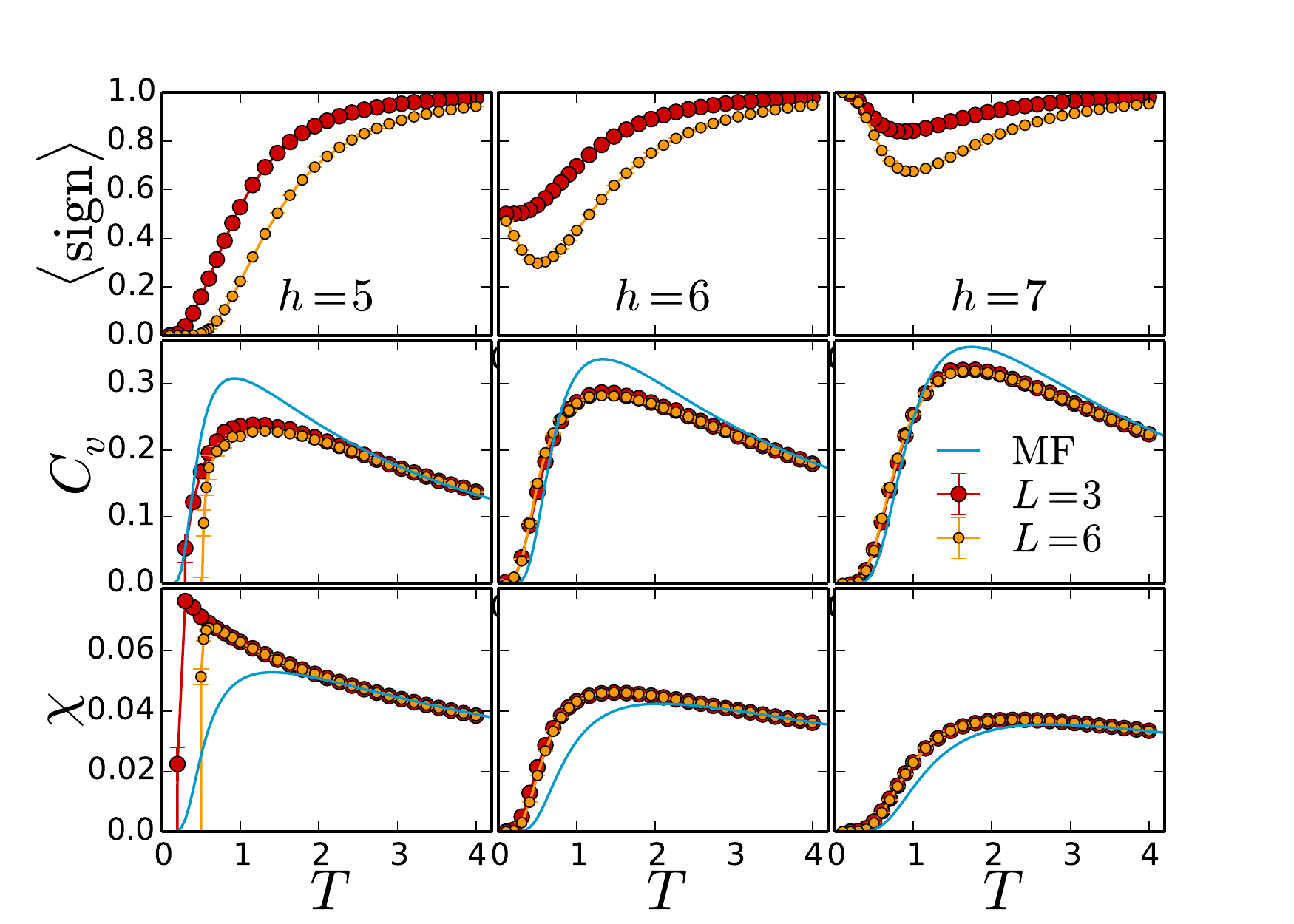}}
\caption{The average sign, specific heat and magnetic susceptibility on small triangular lattice Heisenberg systems as compared with the mean field prediction near the upper saturation threshold $h^{\texttt{+}}_s=6$.  We see that the mean field prediction agrees very well away from the peaks in our measurements, and the strongest deviations are found when the sign goes to zero and efficient simulations are not longer possible.  Finite size effects are invisible even on these small systems except when the sign is nearly zero.}
\label{fig:ThreshTri}
\end{figure}

We have already seen that these mean field results compare extremely well with the data obtained in very high external fields.  We would now like to make the comparison with fields close to $h^{\texttt{+}}_s$, where an easing of the sign first begins to appear at low temperatures.  Fig. \ref{fig:ThreshTri} shows the average sign, specific heat, and magnetic susceptibility obtained on two small triangular lattice systems as compared with the mean field prediction.  We have chosen small lattices here in order to observe the presence of finite size effects.  We find good agreement with mean field theory away from the peaks of our measurements and, as expected, deviations are largest where the sign tends toward zero.  Finite size effects are again essentially absent, except below the threshold at low temperatures where the sign is nearly zero.

\begin{figure}[h!]
\centerline{\includegraphics[angle=0,width=1.0\columnwidth]{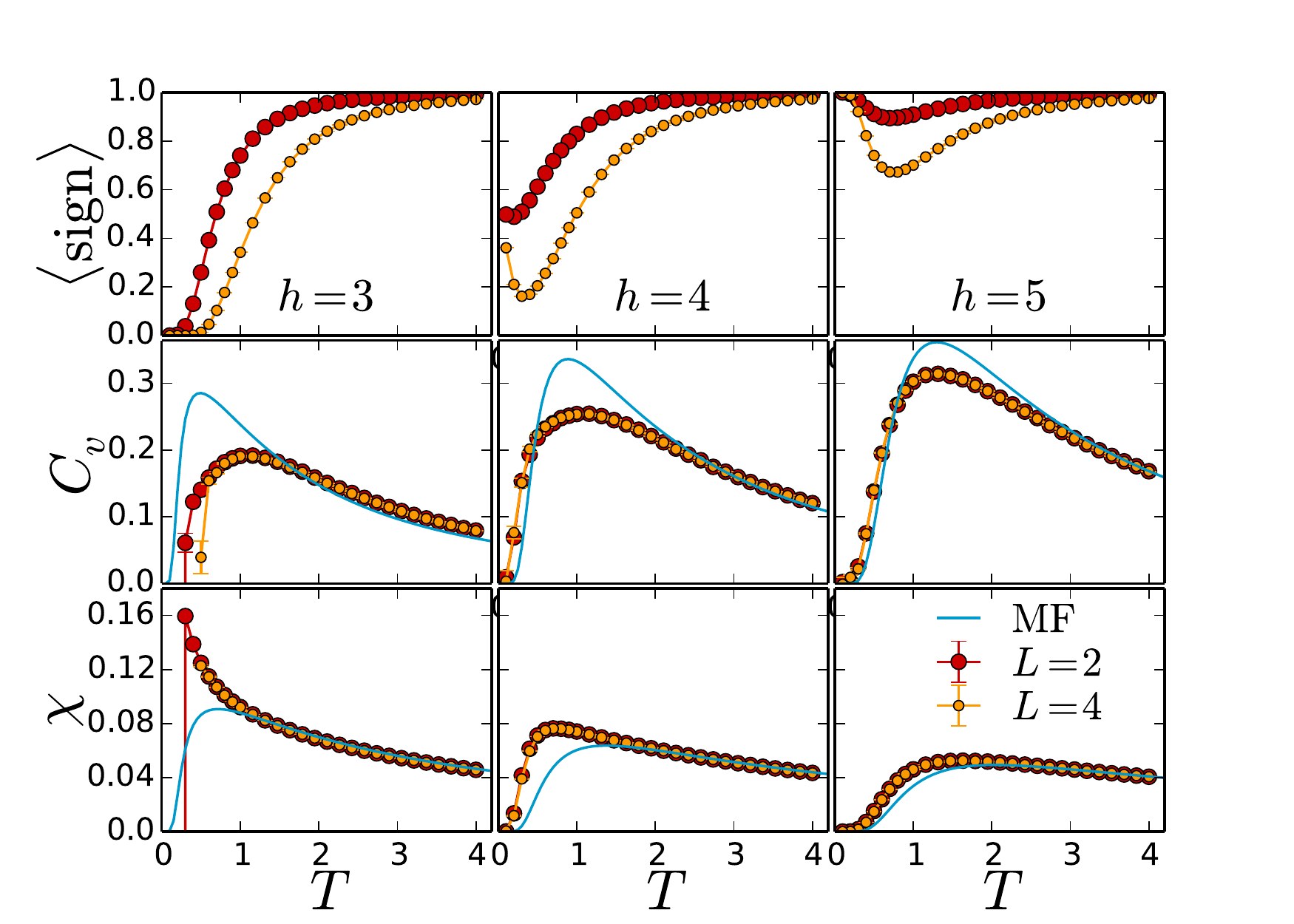}}
\caption{This figure is similar to Fig. \ref{fig:ThreshTri}, except on the kagome lattice where the saturation thresholds are $h^{\texttt{-}}_s=3$ and $h^{\texttt{+}}_s=4$ for $H^{\texttt{-}}$ and $H^{\texttt{+}}$, respectively.  We find a similar agreement with the mean field prediction away from the peaks in $C_v$ and $\chi$, with better agreement at higher fields.  Finite size effects on these small system sizes are equally absent except where the sign is nearly zero.}
\label{fig:ThreshKag}
\end{figure}

In Fig. \ref{fig:ThreshKag} we have performed the same type of comparision, but this time for the kagome lattice.  Here the saturation thresholds for $H^{\texttt{-}}$ and $H^{\texttt{+}}$ are $h^{\texttt{-}}_s=3$ and $h^{\texttt{+}}_s=4$, respectively.  Here we observe a similar absence of finite size effects (except where the sign is nearly zero), and agreement with mean field away from the peaks in $C_v$ and $\chi$.

\section{\label{sec:Dimer}Husimi Heisenberg model in dimer basis}

We now move to a more sophisticated example of the easing of the average sign at low temperatures.  Here we will be interested in ground states that are direct products of singlets, and we will choose the computational basis of dimers (singlet and triplet) for each pair of sites.  Since the ``all singlets" product state is a member of the computational basis, we will find an easing of the sign in the limit when the antiferromagnetic intra-dimer coupling becomes large.

\begin{figure}[h!]
\centerline{\includegraphics[angle=0,width=0.7\columnwidth]{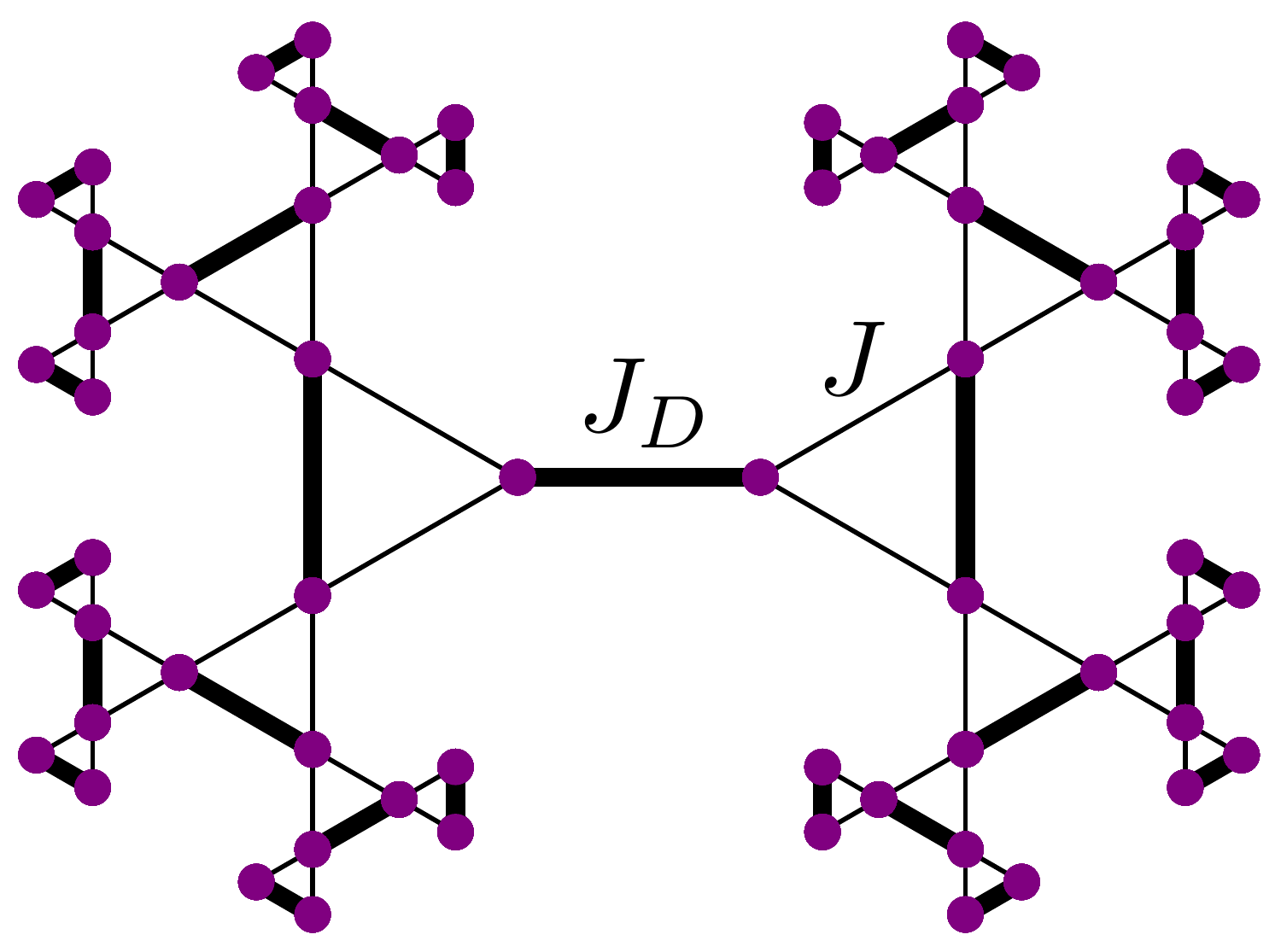}}
\caption{A modified Husimi cactus ($N_{\text{leaf}}=4$, $N_{\text{site}}=62$), which is centered on a bond instead of a site.  The antiferromagnetic Heisenberg model on this lattice has an exact product ground state of singlets on the dimers (thick bonds) when all bonds have equal strength.  The energy gap above this ground state can be increased by introducing intra-dimer couplings ($J_D$) and inter-dimer couplings ($J$) with $J_D/J > 1$.  This allows us to tame the average sign at intermediate temperatures.}
\label{fig:Nleaf4Latt}
\end{figure}

We choose to work with a slightly modified Husimi cactus (see Fig. \ref{fig:Nleaf4Latt}) such that the exact ground state of the Heisenberg antiferromagnet is a product of singlets.  The Husimi cactus has been widely used as a means to approach the physics of the kagome lattice antiferromagnet, whose structure it locally mimics [\onlinecite{Zeng1990:KagomeNet, Elser1993:KagomeHyperbolic, Chandra1994:Husimi, Hao2009:FermionicExcitations, Hao2013:KagLattSL, Liu2014:FeaturelessHusimi, Liao2016:HeisHusimi, Wan2016:PhenomHyperKagome}].

Since the ground state is an exact product state, the goal will be to study the thermodynamic behavior at finite temperatures.  We have also introduced an intra-dimer ($J_D$) and inter-dimer coupling ($J$), which will further allow us to control the average sign.  The Heisenberg Hamiltonian is then simply written as:

\begin{equation}
\label{eq:spinHam}
H = J_D \sum_{( i, j ) \in \, \thickbond} \vec{S}_i \cdot \vec{S}_j + J \sum_{( i, j ) \in \, \thinbond} \vec{S}_i \cdot \vec{S}_j .
\end{equation}

\begin{figure}[t!]
\centerline{\includegraphics[angle=0,width=1.0\columnwidth]{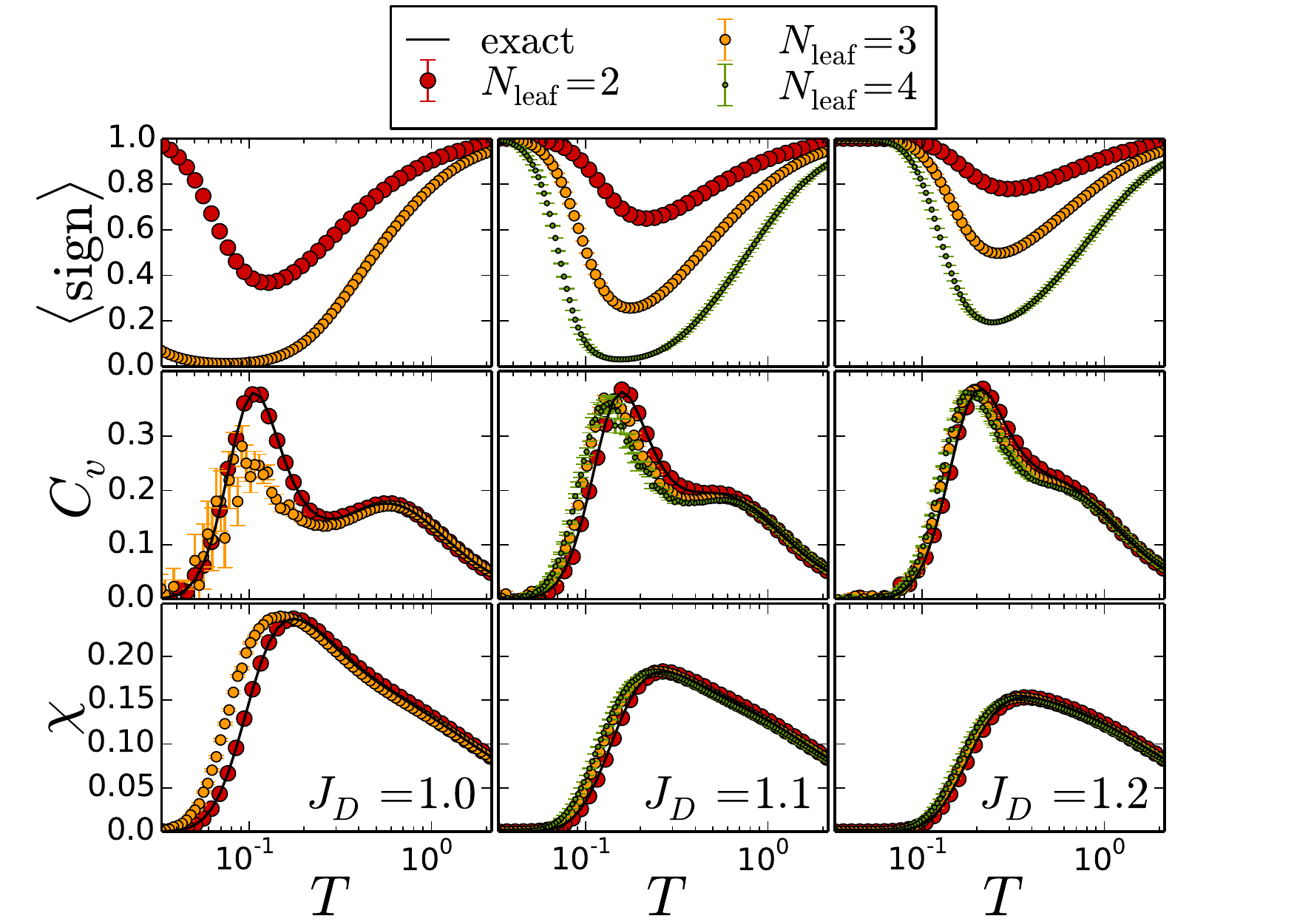}}
\caption{The average sign, specific heat, and magnetic susceptibility for the Husimi lattice antiferromagnet with the inter-dimer coupling $J=1$.  Here we use clusters of size $N_{\mathrm{leaf}}$ ($N_{\mathrm{site}}$) = 2(14), 3(30), and 4(62) in big red, medium orange, and small green markers, respectively.  The black line is obtained by exact diagonalization for $N_{\mathrm{leaf}}=2$, showing perfect agreement with the QMC.  In each case the sign returns to one at low $T$, though the severity of the sign problem at intermediate temperatures for $J_D=1$ prohibits the largest cluster.  Interestingly, we see the appearance of two distinct features in the specific heat (see main text).}
\label{fig:Husimi}
\end{figure}

It is worth noting the zero temperature phase diagram of this model.  As we have mentioned, the ground state of this model with $J_D/J \geq 1$ is a product state of dimer singlets.  In fact, this phase persists on the finite clusters that we have studied to $J_D/J < 1$ as well.  With exact diagonalization we have verified on clusters with $N_{\text{leaf}}=1$ and $N_{\text{leaf}}=2$ that the phase boundary is $J_D/J \approx 0.767$ and $J_D/J \approx 0.857$, respectively.  This can be compared to the same phase boundary in the sign-problem-free model (in the dimer basis), which occurs at $J_D/J \approx 0.829$ and $J_D/J \approx 0.949$ for $N_{\text{leaf}}=1$ and $N_{\text{leaf}}=2$, respectively.

In the limit $J_D/J \to 0$ this model reduces to the Heisenberg antiferromagnet on a bond centered Cayley tree (with the center bond missing).  This is a bipartite lattice, and in this limit we expect long-range antiferromagnetic order in the ground state [\onlinecite{Otsuka1996:DMRGXXZBethe, Friedman1997:DMRGCayley, Kumar2012:DMRGBethe, Changlani2013:HeisAFCayley, Li2012:EfficientBethe}].  To the best of our knowledge the phase diagram at intermediate values of $J_D/J$ is unknown, and unfortunately the present method does not allow us to address this question. In what follows we will consider only the regime where $J_D/J \geq 1$.

The Hamiltonian can be expressed in the dimer basis as we detail in Appendix \ref{appendix:DimerHam}.  A phase of $-1$ is also given to $S^z=0$ triplets on one sublattice to render the minimum number of positive off-diagonal matrix elements in $H^{\texttt{-}}$.  Due to the laboriousness of solving the directed loop equations by hand, we have implemented the linear programming technique suggested in [\onlinecite{Alet2005:GenDirLoop}] that automatically determines all possible loop updating moves and probabilities given the local Hamiltonian matrix elements as input.  We provide a detailed explanation of this approach in Appendix \ref{appendix:LinearProg}.  Additionally, because the Hamiltonian conserves the total spin on each dimer at the edge of the lattice, we find that efficient sampling of the total spin quantum numbers requires parallel tempering [\onlinecite{Marinari1992:SimulatedTemp, Hukushima1996:ExchangeMC}].

We now demonstrate the easing of the sign in the dimer basis for the modified Husimi lattice in Fig. \ref{fig:Husimi}.  Here we consider several lattice sizes, labeled by $N_{\mathrm{leaf}}$, which specifies how many generations of leaves have been added to the central dimer ($N_{\mathrm{dimer}} = 2^{N_{\mathrm{leaf}}+1}-1$).  We have made measurements of the specific heat and magnetic susceptibility as a function of temperature for different values of the intra-dimer coupling $J_D$ (setting $J=1$).  As expected, we find the sign going back to one as $T\to 0$ in all cases.  We note, however, that the low temperature region for efficient simulations may coincide with the region where $C_v$ and $\chi$ are approximately zero.  This is the case when the sign problem at intermediate temperatures is most severe, at $J_D=1$.  Perhaps more importantly than the sign returning to one at low $T$ is the fact that the sign can be controlled by increasing the value of $J_D$.  This allows for the simulation of larger lattices.  However, finite size effects are also smaller in this limit.

Interestingly, we find the presence of a double peak structure in the specific heat, which becomes more pronounced as $J_D$ approaches $J$ from above.  This structure persists on larger lattices, ruling out the possibility of a finite size effect.  The same features in the specific heat have been previously observed in small Husimi cactus clusters [\onlinecite{Zeng1990:KagomeNet}] as well as on the infinite Husimi lattice [\onlinecite{Liu2014:FeaturelessHusimi}].  Similarly for the kagome lattice, various studies have observed two distinct maxima [\onlinecite{Elser1989:NuclearAntiferro, Drzewinski1996:CvKagRG, Elstner1994:CvKagHTE, Nakamura1995:ThermoKag, Tomczak1996:ThermoHeisKag, Khatami2012:ThermoClino, Syromyatnikov2004:LowECvKag, Shimokawa2016:FiniteTempCrossKag, Misguish2005:CvKagHTE}] or a pronounced shoulder feature [\onlinecite{Misguish2005:CvKagHTE, Sugiura2013:CanonThermPureQState, Schnack2018:MagKag, Chen2018:ThermoKag}], which can be attributed to the presence of many low-lying singlet states [\onlinecite{Mila1998:LowEnergySector}].

\section{\label{sec:conc}Conclusions}

We have demonstrated, in two specific cases involving frustrated Heisenberg antiferromagnets, the reduction of the sign problem in QMC simulations at low temperatures when the original signed and the sign-problem-free Hamiltonians have the same ground state and this state is a member of the computational basis.  As a first illustrative example of this, we considered the Heisenberg antiferromagnet on the triangular and kagome lattices in a large external field.  Although this example simply demonstrates the general principle, it is not clear to what extent it could be used to illuminate the thermodynamics of frustrated antiferromagnets.  In fields well above saturation, where efficient Monte Carlo simulations become possible, we found good agreement with the mean field prediction and an almost complete absence of any finite size effects.

In the second example involving the Heisenberg antiferromagnet on the modified Husimi cactus in the dimer basis, the results seem more promising.  Firstly, in the standard $S^z$ basis, the sign problem would be much stronger, most likely preventing any of the results that we have obtained.  More importantly, although the sign returns to one away from the prominent features in $C_v$ and $\chi$, we have a means of controlling the magnitude of the sign by slightly favoring the intra-dimer coupling.  This allows us to ensure efficient simulations, while still being able to observe interesting features in our physical observables, including two broad peaks in the specific heat.

Possible extensions of this work would involve finding other more exotic ground states that could be incorporated into a computational basis, such as the AKLT state [\onlinecite{Affleck1987:AKLT}].  Ideally this basis should be chosen such that it not only satisfies the general criteria that we have outlined, but also so that the positive off-diagonal elements of the Hamiltonian are minimized [\onlinecite{Hangleiter2019:EasingMCSign}].  This would ensure that the sign problem at intermediate temperatures would be as mild as possible.  

{\em Acknowledgements:}
Our QMC simulations were performed on the Fidis cluster at EPFL.

\bibliographystyle{apsrev}
\bibliography{signprobrefs}

\appendix

\section{Husimi Dimer Hamiltonian}
\label{appendix:DimerHam}

Here we will give details on expressing the Hamiltonian in Eq. (\ref{eq:spinHam}) in the dimer basis.  Firstly we can express the Hamiltonian bond operator in terms of the total $\vec{T}_i=\vec{S}_{ia}+\vec{S}_{ib}$ and difference $\vec{D}_i=\vec{S}_{ia}-\vec{S}_{ib}$ operators:

\begin{multline}
\label{eq:Hdimer}
H_{ij} = \tfrac{J_D}{2 n_i} (\vec{T}_i \cdot \vec{T}_i - \tfrac{3}{2}) + \tfrac{J_D}{2 n_j} (\vec{T}_j \cdot \vec{T}_j - \tfrac{3}{2}) +\\ \tfrac{J}{2}  \vec{T}_i \cdot (  \vec{T}_j  \pm   \vec{D}_j ) -(\tfrac{J}{2} + \tfrac{J_D}{4 n_i} + \tfrac{J_D}{4 n_j} + \Delta).
\end{multline}
We indicate that there are two types of bond operators that differ in off-diagonal signs depending on the dimer orientation.  One corresponds to taking $ +   \vec{D}_j$, and the other to $ -   \vec{D}_j$.  Furthermore, since all Hamiltonian terms have been lumped into bond operators, the single dimer terms depend on the coordination number for that dimer ($n_i$ and $n_j$ for dimer $i$ and dimer $j$).  We have also subtracted a constant to render all of the diagonal matrix elements negative or zero, and $\Delta \geq 0$ can be used to further shift the matrix elements for efficient QMC sampling (here we've taken $\Delta$=0.2 in our simulations).

\begin{figure}[t!]
\centerline{\includegraphics[angle=0,width=0.7\columnwidth]{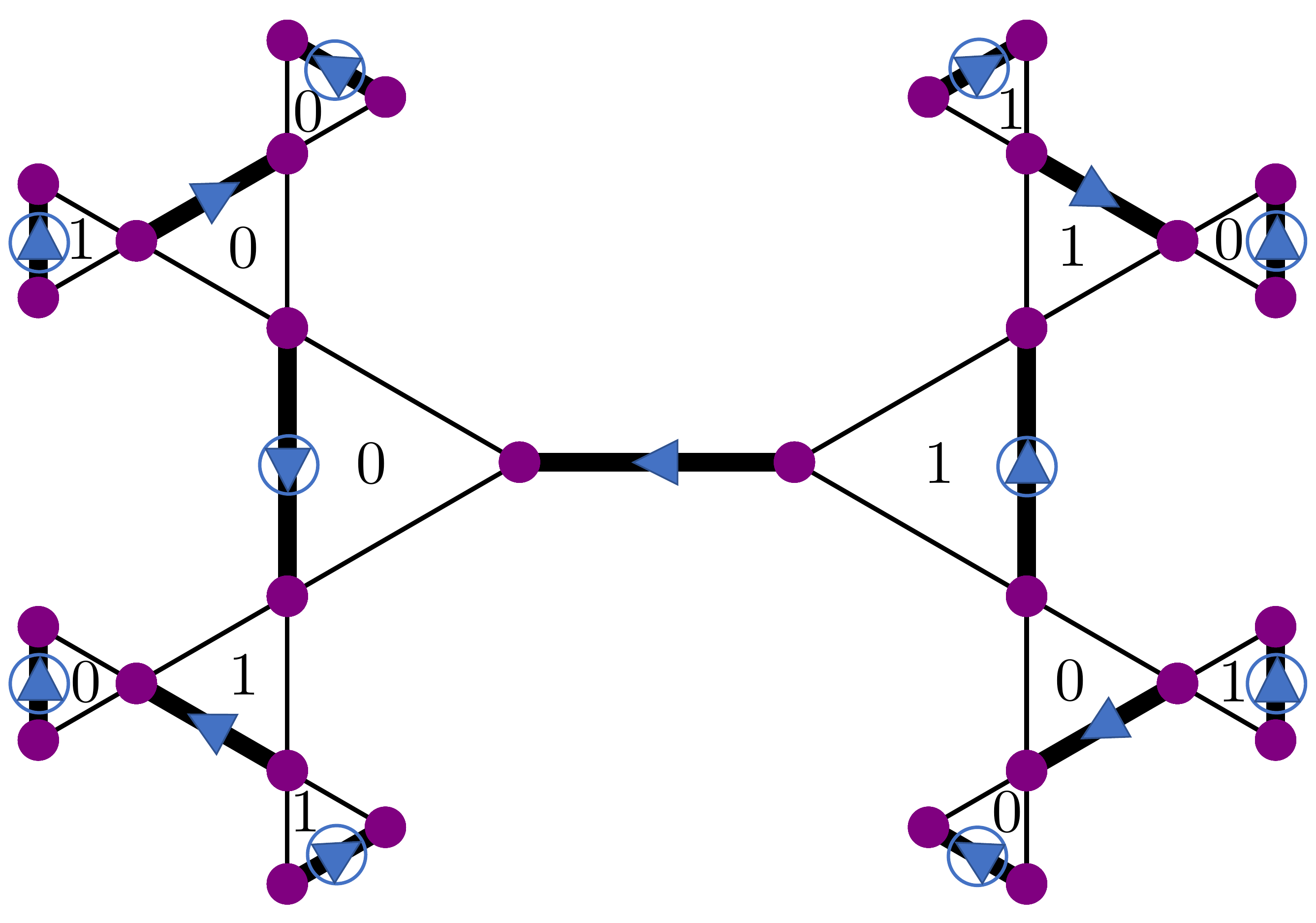}}
\caption{Here, on an $N_{\text{leaf}}=3$ system, we indicate the orientation of the singlets (arrows) and $-1$ phases of the $S^z=0$ triplet (blue circles) that have been chosen for the dimers.  This results in an off-diagonal sign structue of either type-0 (Eq (\ref{eq:Hodiag0})) or type-1 (Eq (\ref{eq:Hodiag1})) in the Hamiltonian bond operator.  The corresponding types for this arrangement are indicated between the dimers.}
\label{fig:husimibasis}
\end{figure}

Once the $\vec{T}_i$ and $\vec{D}_i$ operators have been expressed in the dimer basis, we find that phases need to be introduced on one sublattice of the (bipartite) binary tree lattice formed by the dimers.  This results in a minimum number of positive off-diagonal (sign-problem causing) matrix elements.  In Fig. \ref{fig:husimibasis} we depict the orientation and phase structure that we have used on the Husimi cactus.  Here arrows represent the orientation of the dimers, and circled arrows mean that the $S^z=0$ triplet on that dimer has been given a phase -1.  This results in Hamiltonian bond operators that have a sign structure labeled by either 0 or 1 (see Eq. (\ref{eq:Hodiag0}) and Eq. (\ref{eq:Hodiag1})).  Notice that arrows always point to type-0 and away from type-1, while circled arrows point to type-1 and away from type-0.

We choose to order the states in the two-dimer Hilbert space as $\{| \bullet \rangle, | 0 \rangle, | + \rangle, | - \rangle\}^{\otimes 2}$.  The diagonal part of the bond Hamiltonian is then expressed as:
\begin{multline}
\label{eq:Hdiag}
H^{\text{d}}_{ij}=-\tfrac{J}{2} - \tfrac{J_D}{n_i} - \tfrac{J_D}{n_j} - \Delta \, +  \\
\text{diag}(
0,
\tfrac{J_D}{n_j},
\tfrac{J_D}{n_j},
\tfrac{J_D}{n_j},
\tfrac{J_D}{n_i},
\tfrac{J_D}{n_i}+\tfrac{J_D}{n_j},
\tfrac{J_D}{n_i}+\tfrac{J_D}{n_j},
\tfrac{J_D}{n_i}+\tfrac{J_D}{n_j},\\
\tfrac{J_D}{n_i},
\tfrac{J_D}{n_i}+\tfrac{J_D}{n_j},
\tfrac{J}{2} + \tfrac{J_D}{n_i} + \tfrac{J_D}{n_j},
-\tfrac{J}{2} + \tfrac{J_D}{n_i} + \tfrac{J_D}{n_j},
\tfrac{J_D}{n_i},\\
\tfrac{J_D}{n_i}+\tfrac{J_D}{n_j},
-\tfrac{J}{2} + \tfrac{J_D}{n_i} + \tfrac{J_D}{n_j},
\tfrac{J}{2} + \tfrac{J_D}{n_i} + \tfrac{J}{n_j}
).
\end{multline}
The off-diagonal part of the Hamiltonian can be expressed in the subspace: $\{| 0 \bullet \rangle, | 0 0 \rangle, | 0 + \rangle, | 0 - \rangle, | + \bullet \rangle, | + 0 \rangle, | + - \rangle, | - \bullet \rangle,  | - 0 \rangle,  | - + \rangle \}$ as
\begin{equation}
\label{eq:Hodiag0}
H^{\text{0}\mp}_{ij}=-\frac{J}{2}
\begin{bmatrix} 
0      & 0 & 0      & 0 & 0      & 0      & 1 & 0 & 0 & \mp1\\
0      & 0 & 0      & 0 & 0      & 0      & 1 & 0 & 0 & 1\\
0      & 0 & 0      & 0 & \mp1& 1      & 0 & 0 & 0 & 0\\
0      & 0 & 0      & 0 & 0      & 0      & 0 & 1 & 1 & 0\\
0      & 0 & \mp1& 0 & 0      & \mp1& 0 & 0 & 0 & 0\\
0      & 0 & 1      & 0 & \mp1& 0      & 0 & 0 & 0 & 0\\
1      & 1 & 0      & 0 & 0      & 0      & 0 & 0 & 0 & 0\\
0      & 0 & 0      & 1 & 0      & 0      & 0 & 0 & 1 & 0\\
0      & 0 & 0      & 1 & 0      & 0      & 0 & 1 & 0 & 0\\
\mp1& 1 & 0      & 0 & 0      & 0      & 0 & 0 & 0 & 0\\
\end{bmatrix}
\end{equation}
corresponding to the sign structure of type-0 in Fig. \ref{fig:husimibasis}, and the type-1 is given by
\begin{equation}
\label{eq:Hodiag1}
H^{1\mp}_{ij}=-\frac{J}{2}
\begin{bmatrix} 
0      & 0 & 0      & 0       & 0      & 0      & \mp1 & 0       & 0       & 1\\
0      & 0 & 0      & 0       & 0      & 0      & 1       & 0       & 0       & 1\\
0      & 0 & 0      & 0       & 1      & 1      & 0       & 0       & 0       & 0\\
0      & 0 & 0      & 0       & 0      & 0      & 0       & \mp1 & 1       & 0\\
0      & 0 & 1      & 0       & 0      & 1      & 0       & 0       & 0       & 0\\
0      & 0 & 1      & 0       & 1      & 0      & 0       & 0       & 0       & 0\\
\mp1& 1 & 0      & 0       & 0      & 0      & 0       & 0       & 0       & 0\\
0      & 0 & 0      & \mp1 & 0      & 0      & 0       & 0       & \mp1 & 0\\
0      & 0 & 0      & 1       & 0      & 0      & 0       & \mp1 & 0       & 0\\
1      & 1 & 0      & 0       & 0      & 0      & 0       & 0       & 0       & 0\\
\end{bmatrix}
\end{equation}
We have also indicated the signed ($-$) and sign-free ($+$) Hamiltonian by using a superscript $\mp$. 

\section{Linear programming QMC}
\label{appendix:LinearProg}

\subsection{Generalities}
\label{appsub:gens}

As we have seen in Appendix \ref{appendix:DimerHam}, the Hamiltonian in the dimer basis is quite complicated.  It is a $16 \times 16$ matrix with all diagonal matrix elements nonzero.  Furthermore, since all of the interactions have been gathered into bond operators, the bond operators depend on the coordination numbers of the sites.

In order to efficiently sample the partition function for this model, one needs to solve the directed loop equations [\onlinecite{Syljuasen2002:DirectedLoops}] associated with each matrix element in the Hamiltonian.  For this to be carried out by hand is overly tedious, and prone to human error.  We have therefore implemented a linear programming technique suggested in [\onlinecite{Alet2005:GenDirLoop}] that, given a Hamiltonian matrix element and a loop operator type, automatically determines the possible update moves and probabilities with a minimal chance of bouncing.  This is computed for every matrix element during the initialization of the simulation, and these probabilities are then used during the loop updates.  We now describe the necessary ingredients for implementing this technique.

\subsection{Worm types}
\label{appsub:worm}

We begin by assuming that we have a loop operator type defined by $T(s)=s^{\texttt{+}}, \, T^{\dagger}(s)=s^{\texttt{-}}$, also referred to as a worm [\onlinecite{Alet2005:GenDirLoop}].  We also have that $T(s^{\texttt{-}})=T^{\dagger}(s^{\texttt{+}})=s$.  This defines a loop type that will be used to update the spin states ($s$) attached to matrix elements in the imaginary time configuration.  The notion of an operator and its conjugate is important, since this defines the transformation that happens to spin states depending on whether the worm is moving up or down in imaginary time.  When the worm is moving up one uses $T$ to transform the spin and when the worm is moving down one uses $T^{\dagger}$ (or vice versa).  As an example, the bounce process corresponds to changing one spin on a matrix element using $T(s)=s^{\texttt{+}}$, then immediately reversing direction (switching to the conjugate operator) and changing the spin back using $T^{\dagger}(s^{\texttt{+}})=s$.  Several worm types may need to be defined in order to make the simulation ergodic.  In practice this means that repeatedly using the various worm types allows for the generation of any matrix element in the Hamiltonian given any starting matrix element.  We have found that our simulations of the Husimi Heisenberg model in the dimer basis are ergodic if we define two worm types:  

\begin{equation}
\label{eq:T1}
T_1=
\begin{bmatrix} 
0 & 0 & 0 & 0\\
0 & 0 & 0 & 1\\
0 & 1 & 0 & 0\\
0 & 0 & 0 & 0\\
\end{bmatrix} \quad\quad
T_2=
\begin{bmatrix} 
0 & 0 & 0 & 0\\
0 & 0 & 0 & 1\\
1 & 0 & 0 & 0\\
0 & 0 & 0 & 0\\
\end{bmatrix}
\end{equation}

Where again the basis is ordered as $\{| \bullet \rangle, | 0 \rangle, | + \rangle, | - \rangle\}$.  $T_1$ simply raises the z-component of the spin in the triplet sector, and $T_2$ converts $| - \rangle \to | 0 \rangle$ and $| \bullet \rangle \to | + \rangle$.  There are other combinations of worm types that are ergodic, and it would be interesting to explore which combinations are most efficient.

\begin{figure}[t!]
\centerline{\includegraphics[angle=0,width=0.8\columnwidth]{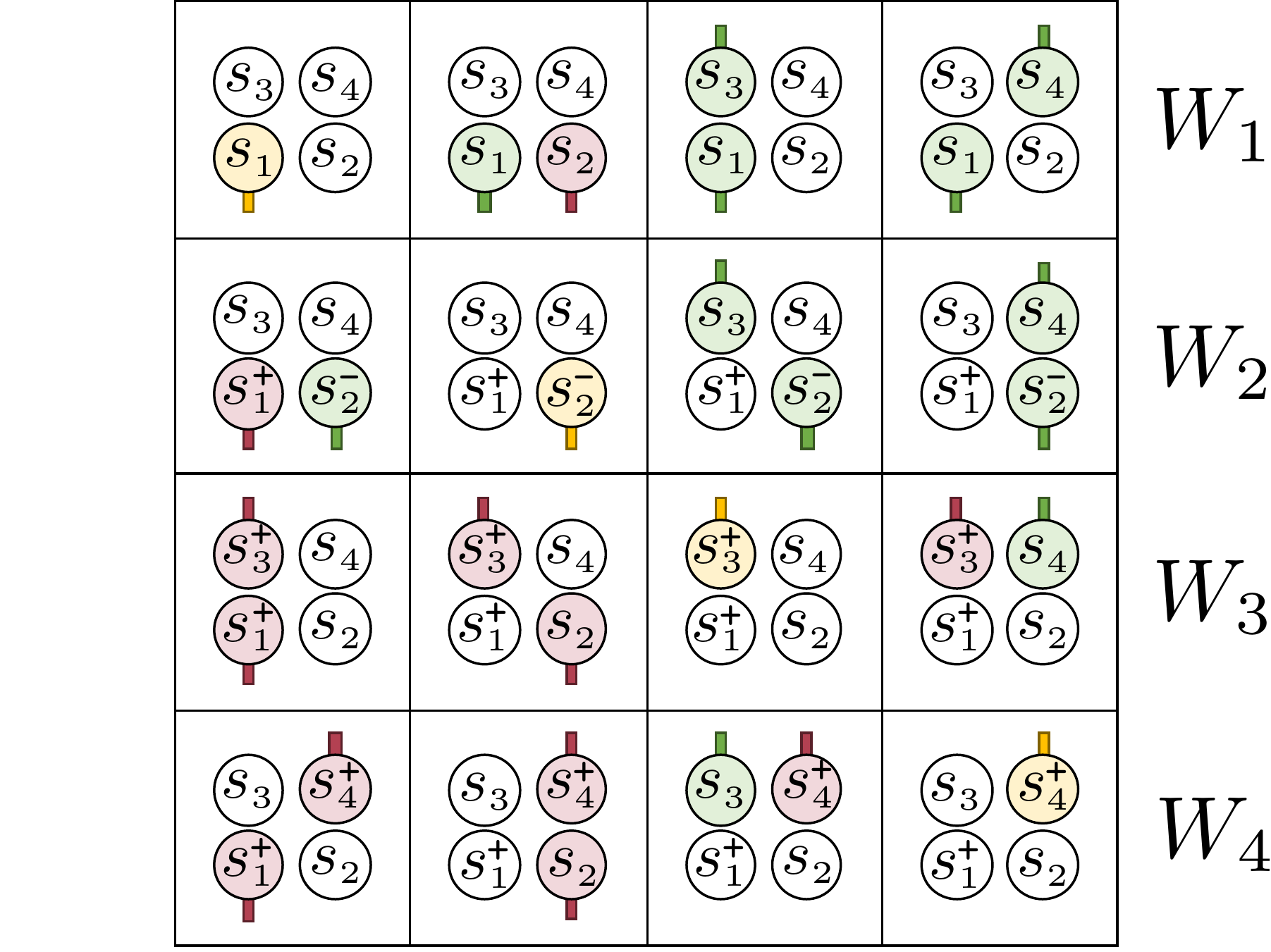}}
\caption{The general set of processes that are required to satisfy detailed balance among themselves.  Here each process and its conjugate are listed ($a_{ij}$ and $a_{ji}$ for $i \neq j$, where $i$ and $j$ label the rows and columns in the table).  The bounce processes are along the diagonal $a_{ii}$.  Here the color refers to the direction of travel for the worm head.  Green means traveling up (transform spin with $T$), red means traveling down (transform spin with $T^{\dagger}$), and yellow means bounce (don't change the spin).  The $W_i$ are the values of the matrix elements appearing in each row (some of which may be zero).}
\label{fig:dle}
\end{figure}

\subsection{Matrix elements and detailed balance}
\label{appsub:mel}

We now consider a specific matrix element of the bondHamiltonian, labeled by spin values $s_1, s_2, s_3, s_4$.  Here $s_1$ and $s_2$ label the incoming bond state and $s_3$ and $s_4$ label the outgoing bond state.  This matrix element is used to generate a table of update processes that together will be required to satisfy detailed balance (Fig. \ref{fig:dle}).  This is done by selecting an incoming spin (for example $s_1$) and a worm type.  In Fig. \ref{fig:dle}, green means the worm is traveling up (use $T$ to transform spin), red means down (use $T^{\dagger}$), and yellow means bounce (don't change the spin).  The $W_i$ label the values of the matrix elements appearing in each row.  Each process can be assigned an amplitude $a_{ij}$, where $ij$ labels the row and column in Fig. \ref{fig:dle}.  The probabilities are defined as $P_{ij}=a_{ij}/W_i$. We then enforce $\sum_j a_{ij}=W_i$ which ensures that the probability for all processes in a row add up to one.  Finally, to satisfy detailed balance we set $a_{ij}=a_{ji}$ [\onlinecite{Syljuasen2002:DirectedLoops}], resulting in only ten free parameters ($a_{11}, a_{12}, a_{13}, a_{14}, a_{22}, a_{23}, a_{24}, a_{33}, a_{34}, a_{44}$). 

\subsection{The simplex tableau}
\label{appsub:mel}

We shall organize all of the information as follows:

\begin{align}
a_{11} &+ a_{12} + a_{13} + a_{14}  = W_1 \label{eq:lineq1}\\
a_{22} &+ a_{12} + a_{23} + a_{24}  = W_2 \label{eq:lineq2}\\
a_{33} &+ a_{13} + a_{23} + a_{34}  = W_3 \label{eq:lineq3}\\
a_{44} &+ a_{14} + a_{24} + a_{34}  = W_4 \label{eq:lineq4}\\
z &-2 \sum_{i < j} a_{ij} =  -\sum_i W_i   \label{eq:lineq5}
\end{align}
where we have introduced the objective function $z = -\sum_i a_{ii}$ that we wish to maximize (the same as minimizing the total bounce probability) and have re-expressed it using the first four equations.  We have organized the equations so that the bounces are listed only in the leftmost column.  As such, the bounces should be regarded as dependent variables and are determined once the independent variables $a_{ij}$ ($i < j$) are specified.

\begin{table}
\begin{center}
 \begin{tabular}{||c !{\vrule width1.0pt} c | c | c | c | c | c | c ||} 
 \hline
  & $a_{12}$ & $a_{13}$ & $a_{14}$ & $a_{23}$ & $a_{24}$ & $a_{34}$ &  \\ [0.5ex] 
\Xhline{2.5\arrayrulewidth}
 $a_{11}$ & 1 & 1 & 1 & 0 & 0 & 0 & $W_1$ \\ 
 \hline
 $a_{22}$ & 1 & 0 & 0 & 1 & 1 & 0 & $W_2$ \\
 \hline
 $a_{33}$ & 0 & 1 & 0 & 1 & 0 & 1 & $W_3$ \\
 \hline
 $a_{44}$ & 0 & 0 & 1 & 0 & 1 & 1 & $W_4$ \\ 
\hline
 $z$ & -2 & -2 & -2 & -2 & -2 & -2 & $-\sum_i W_i$ \\  
 \hline
\end{tabular}
\end{center}
\caption{The simplex tableau in canonical form, with the bounces listed to the left as dependent variables.  Setting the independent variables to zero $a_{ij}=0$ $(i<j)$  (in the top row) gives the initial feasible solution, which is the least optimal.  In this solution the bounce amplitudes are given by the values in the last column and the bounce probabilities are all maximal.  The ``-2" values in the last row are the negative of the coefficients in the objective function, and the bottom right corner gives the value of the objective function evaluated at the solution.}
\label{tab:linprog2}
\end{table}

In Table (\ref{tab:linprog2}) we have arranged the same information into a simplex tableau [\onlinecite{Dantzig1963:LinProg, Ferguson:LP}].  The variables appearing on the top row are regarded as independent, whereas the variables in the left column are dependent.  The values ``-2" appearing in the last row are the negative of the coefficients in the objective function $z$, and $-\sum_i W_i$ is the value of the objective function when the independent variables are set to zero.  

We see that setting the independent variables equal to zero constitutes a solution to the tableau (solving the directed loop equations).  In this solution, the bounce amplitudes ($a_{ii}$) are determined by the rightmost column ($a_{ii}=W_i$ and $P_{ii}=1$).  Though this is a feasible solution, it is the least optimal.  We can therefore look for another feasible solution that increases the value of the objective function.  It is clear that the objective function can be increased because the values in the last row are negative.  This means that the coefficients in the objective function are positive, so increasing the independent variables increases the function.  Conversely, if we had all positive entries in the bottom row (except the bottom right corner), we would find an optimal solution by setting the independent variables equal to zero.  This would mean that the coefficients of the objective function are negative, and so the best we can do to maximize it is to set the independent variables to zero.

The simplex algorithm [\onlinecite{Dantzig1963:LinProg, Ferguson:LP}] (this appendix follows Ref. [\onlinecite{Ferguson:LP}]) allows us to cycle through the possible solutions, while systematically approaching an optimal solution.  This is done by exchanging dependent variables with independent variables via the pivot operation.  Once one has pivoted such that all the ``-2" values in the last row become numbers greater than or equal to zero, an optimal solution is found.  The amplitudes in the top row are zero and the amplitudes in the left column are given by the values in the right column.

\subsection{Pivoting}
\label{appsub:pivoting}

We now describe the process of exchanging a dependent variable for an independent variable, which is referred to as pivoting.  This corresponds to solving one equation for an independent variable (say $a_{12}$) and using this equation to replace the occurrences of $a_{12}$ in all other equations, including the objective function.

Two things happen to the simplex tableau during a pivot operation.  First, the dependent and independent variables trade places in the tableau.  So if we chose to exchange $a_{12}$ with $a_{11}$, then $a_{12}$ would be written in the left column in the original location of $a_{11}$ and similarly $a_{11}$ would be written on the top row in the original location of $a_{12}$.

The other thing that happens during a pivot is that the values in the tableau will change.  This includes the weights in the right column, the negative coefficients of the objective function in the bottom row, and the value of the objective function evaluated at the solution (bottom right corner).

There is a simple set of rules to describe how these values change after a pivot, which is summarized in Table (\ref{tab:pivot}).  Here $p$ is the pivot element itself, which is in the row of a dependent variable and a column of an independent variable.  After pivoting this element goes into the reciprocal of itself.  All elements in the same row (except the pivot element) get divided by the pivot.  All elements in the same column (except the pivot element) get divided by the pivot and multiplied by $-1$.  And finally, all elements $q$ that are neither in the same row nor the same column change to $q-(rc)/p$, were $r$ is the element that shares the same row as the pivot (above or below $q$) and $c$ is the element that shares the same column as the pivot (to the left or right of $q$).  Again, these rules are to be applied to all of the values appearing in the tableau (excluding the cells that label the names of the variables).

\begin{table}[h!]
    \centering \quad \quad \quad \quad \, \,
    \begin{tabular}[t]{| c | c |}
        \hline
        $p$ & $r$      \\
        \hline
        $c$ & $q$      \\
        \hline
    \end{tabular} \quad
    \begin{tabular}[t]{ c }
                     \\
        $\myarrow$      \\ 
    \end{tabular} \quad
    \begin{tabular}[t]{| c | c |}
        \hline
        $1/p$ & $r/p$      \\ 
        \hline
        $-c/p$ & $q-(rc)/p$      \\
        \hline
    \end{tabular} \quad \quad \quad \quad \quad
    \caption{A summary of the rules for pivoting the simplex tableau.  The pivot element goes into it's reciprocal, the pivot row (except the pivot) gets divided by the pivot, the pivot column (except the pivot) gets divided by the pivot and multiplied by -1, and all other elements $q$ get reduced by $q \to q-(rc)/p$.  Here $r$ is the element that shares the same row as the pivot and the same column as $q$, and $c$ shares the same column as the pivot and the same row as $q$.}
    \label{tab:pivot}
\end{table}

\subsection{The simplex algorithm}
\label{appsub:simpalg}

We are now have all of the ingredients to perform the simplex algorithm to systematically approach an optimal solution.  This tells us how to choose our pivots so as to improve the solution.  The rule is the following: take any column $j$ (corresponding to an independent variable) where the value in the last row is negative, this is the pivot column.  Next consider all the possible pivots $p_{ij} > 0$ in column $j$ (corresponding to different dependent variables).  The one with the smallest value $\tilde{W}_i/p_{ij}$ is chosen as the pivot, and if there is a tie any one can be chosen. Here $\tilde{W}_i$ designates the current value in the last column of row $i$.  This procedure is iterated until all the values in the last row (except the bottom right corner) are positive or zero.  When an optimal solution is found, the independent variables are zero and the dependent variables take on the values in the last column.

We can notice some properties about this algorithm.  Firstly, all of the weights $\tilde{W}_i$ remain positive.  This comes from always choosing the pivot with the smallest value $\tilde{W}_i/p_{ij}$.  Also, the value of the objective function in the bottom right corner will alway increase or stay the same.  This comes from the fact that $\tilde{W}_i$ is always positive (or zero) and we always choose a pivot column with the last entry negative.  We are guaranteed to be able to apply the rules of the simplex algorithm.  For example, one will not encounter a situation where the pivot column contains only negative elements.  If this is the case then the problem is unbounded feasible, which will not occur here.

\begin{figure}[h!]
\centerline{\includegraphics[angle=0,width=0.8\columnwidth]{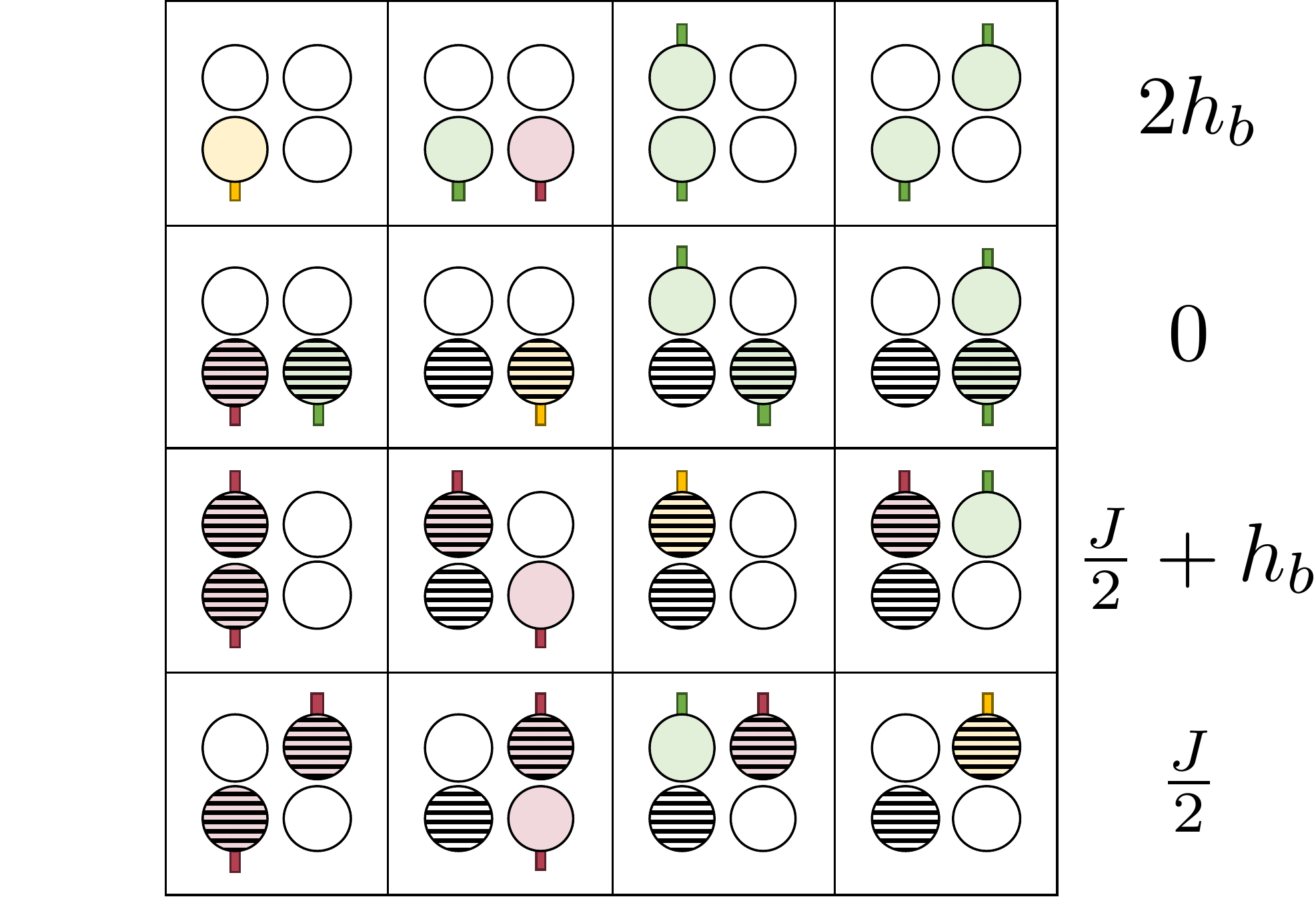}}
\caption{The directed loop moves for a set of matrix elements of the Heisenberg model in a uniform magnetic field.  Unshaded (shaded) circles mean spin up (down), and the color represents the travel direction for the worm.  Green is traveling up, red is down and yellow is bounce.  The weights on the right are the values of the matrix elements in each row.}
\label{fig:dleEX}
\end{figure}

\begin{table}

\caption{The initial tableau in canonical form.  The first pivot is chosen at (2,1).}\label{tab:EX1}
 \begin{tabular}{||c !{\vrule width1.0pt} c | c | c | c | c | c | c ||} 
 \hline
  & $a_{12}$ & $a_{13}$ & $a_{14}$ & $a_{23}$ & $a_{24}$ & $a_{34}$ &  \\ [0.5ex] 
\Xhline{2.5\arrayrulewidth}
 $a_{11}$ & 1 & 1 & 1 & 0 & 0 & 0 & $2 h_b$ \\ 
 \hline
 $a_{22}$ & \textcircled{1} & 0 & 0 & 1 & 1 & 0 & $0$ \\
 \hline
 $a_{33}$ & 0 & 1 & 0 & 1 & 0 & 1 & $\tfrac{J}{2}+h_b$ \\
 \hline
 $a_{44}$ & 0 & 0 & 1 & 0 & 1 & 1 & $\tfrac{J}{2}$ \\ 
\hline
 $z$ & -2 & -2 & -2 & -2 & -2 & -2 & $-(J+3h_b)$ \\  
 \hline
\end{tabular}

\bigskip

\caption{After the first pivot.  The next pivot is (1,2).}\label{tab:EX2}

 \begin{tabular}{||c !{\vrule width1.0pt} c | c | c | c | c | c | c ||} 
 \hline
  & $a_{22}$ & $a_{13}$ & $a_{14}$ & $a_{23}$ & $a_{24}$ & $a_{34}$ &  \\ [0.5ex] 
\Xhline{2.5\arrayrulewidth}
 $a_{11}$ & -1 & \textcircled{1} & 1 & -1 & -1 & 0 & $2 h_b$ \\ 
 \hline
 $a_{12}$ & 1 & 0 & 0 & 1 & 1 & 0 & $0$ \\
 \hline
 $a_{33}$ & 0 & 1 & 0 & 1 & 0 & 1 & $\tfrac{J}{2}+h_b$ \\
 \hline
 $a_{44}$ & 0 & 0 & 1 & 0 & 1 & 1 & $\tfrac{J}{2}$ \\ 
\hline
 $z$ & 2 & -2 & -2 & 0 & 0 & -2 & $-(J+3h_b)$ \\  
 \hline
\end{tabular}

\bigskip

\caption{After the second pivot.  The next pivot is (2,4).}\label{tab:EX3}

\begin{tabular}{||c !{\vrule width1.0pt} c | c | c | c | c | c | c ||} 
 \hline
  & $a_{22}$ & $a_{11}$ & $a_{14}$ & $a_{23}$ & $a_{24}$ & $a_{34}$ &  \\ [0.5ex] 
\Xhline{2.5\arrayrulewidth}
 $a_{13}$ & -1 & 1 & 1 & -1 & -1 & 0 & $2 h_b$ \\ 
 \hline
 $a_{12}$ & 1 & 0 & 0 & \textcircled{1} & 1 & 0 & $0$ \\
 \hline
 $a_{33}$ & 1 & -1 & -1 & 2 & 1 & 1 & $\tfrac{J}{2}-h_b$ \\
 \hline
 $a_{44}$ & 0 & 0 & 1 & 0 & 1 & 1 & $\tfrac{J}{2}$ \\ 
\hline
 $z$ & 0 & 2 & 0 & -2 & -2 & -2 & $-(J-h_b)$ \\  
 \hline
\end{tabular}

\bigskip

\caption{After the third pivot.  The next pivot is (3,6).}\label{tab:EX4}

\begin{tabular}{||c !{\vrule width1.0pt} c | c | c | c | c | c | c ||} 
 \hline
  & $a_{22}$ & $a_{11}$ & $a_{14}$ & $a_{12}$ & $a_{24}$ & $a_{34}$ &  \\ [0.5ex] 
\Xhline{2.5\arrayrulewidth}
 $a_{13}$ & 0 & 1 & 1 & 1 & 0 & 0 & $2 h_b$ \\ 
 \hline
 $a_{23}$ & 1 & 0 & 0 & 1 & 1 & 0 & $0$ \\
 \hline
 $a_{33}$ & -1 & -1 & -1 & -2 & -1 & \textcircled{1} & $\tfrac{J}{2}-h_b$ \\
 \hline
 $a_{44}$ & 0 & 0 & 1 & 0 & 1 & 1 & $\tfrac{J}{2}$ \\ 
\hline
 $z$ & 2 & 2 & 0 & 2 & 0 & -2 & $-(J-h_b)$ \\  
 \hline
\end{tabular}

\bigskip

\caption{After the fourth pivot.  The next pivot is (4,3).}\label{tab:EX5}

\begin{tabular}{||c !{\vrule width1.0pt} c | c | c | c | c | c | c ||} 
 \hline
  & $a_{22}$ & $a_{11}$ & $a_{14}$ & $a_{12}$ & $a_{24}$ & $a_{33}$ &  \\ [0.5ex] 
\Xhline{2.5\arrayrulewidth}
 $a_{13}$ & 0 & 1 & 1 & 1 & 0 & 0 & $2 h_b$ \\ 
 \hline
 $a_{23}$ & 1 & 0 & 0 & 1 & 1 & 0 & $0$ \\
 \hline
 $a_{34}$ & -1 & -1 & -1 & -2 & -1 & 1 & $\tfrac{J}{2}-h_b$ \\
 \hline
 $a_{44}$ & 1 & 1 & \textcircled{2} & 2 & 2 & -1 & $h_b$ \\ 
\hline
 $z$ & 0 & 0 & -2 & -2 & -2 & 2 & $-h_b$ \\  
 \hline
\end{tabular}

\bigskip

\caption{The final tableau with an optimal solution.}\label{tab:EX6}

\begin{tabular}{||c !{\vrule width1.0pt} c | c | c | c | c | c | c ||} 
 \hline
  & $a_{22}$ & $a_{11}$ & $a_{44}$ & $a_{12}$ & $a_{24}$ & $a_{33}$ &  \\ [0.5ex] 
\Xhline{2.5\arrayrulewidth}
 $a_{13}$ & -$\tfrac{1}{2}$ & $\tfrac{1}{2}$ & -$\tfrac{1}{2}$ & 0 & -1 & $\tfrac{1}{2}$ & $\tfrac{3 h_b}{2}$ \\ 
 \hline
 $a_{23}$ & 1 & 0 & 0 & 1 & 1 & 0 & $0$ \\
 \hline
 $a_{34}$ & -$\tfrac{1}{2}$ & -$\tfrac{1}{2}$ & $\tfrac{1}{2}$ & -1 & 0 & $\tfrac{1}{2}$ & $\tfrac{J}{2}-\tfrac{h_b}{2}$ \\
 \hline
 $a_{14}$ & $\tfrac{1}{2}$ & $\tfrac{1}{2}$ & $\tfrac{1}{2}$ & 1 & 1 & -$\tfrac{1}{2}$ & $\tfrac{h_b}{2}$ \\ 
\hline
 $z$ & 1 & 1 & 1 & 0 & 0 & 1 & $0$ \\  
 \hline
\end{tabular}

\end{table}

\subsection{Example problem}
\label{appsub:example}

We would like to illustrate the simplex algorithm in action by providing an explicit example.  We will consider a set of matrix elements and update moves that appears in the Heisenberg antiferromagnet in a magnetic field.

\begin{equation}
\label{eq:Hmag}
H = \sum_{\langle i j \rangle} \left(J \vec{S}_i \cdot \vec{S}_j  - h_b(S^z_i + S^z_j)\right).
\end{equation}

Here the $h_b=h/N_c$ (the external field divided by the coordination number).  On a bipartite lattice the Hamiltonian is sign-free, and by a constant shift of $-(\tfrac{J}{4}+h_b)$ and a sublattice rotation, can be written as

\begin{equation}
\label{eq:Hmagmat}
H_{ij} =-
\begin{bmatrix} 
2 h_b & 0 & 0 & 0\\
0 & \tfrac{J}{2}+h_b & \tfrac{J}{2} & 0\\
0 & \tfrac{J}{2} & \tfrac{J}{2}+h_b & 0\\
0 & 0 & 0 & 0\\
\end{bmatrix}.
\end{equation}

One then needs to consider the table of update moves in Fig. \ref{fig:dleEX}, where the up spin is denoted as an unshaded circle and the down spin is shaded.  Again the colors indicate the direction of the traveling worm head (green is up, red is down and yellow is a bounce).  The worm type in this case is chosen as the $\sigma^x$ operator ($T=T^{\dagger}=\sigma^x$).

To determine the update amplitudes $a_{ij}$ of the processes contained in the table, we initialize the simplex tableau as in Table \ref{tab:EX1}.  Here we have circled the first pivot, and in the following pivots we assume $h_b \ll J$.  In this limit an optimal solution is thus given by $a_{13}=\tfrac{3 h_b}{2}, a_{14}=\tfrac{h_b}{2}, a_{34}=\tfrac{J}{2}-\tfrac{h_b}{2}$, with the other amplitudes zero.  Here the bounces can be completely excluded, as seen by the value of the objective function in the final tableau.

If instead one considers the case $h_b \gg J$, a different sequence of pivots, given by (2,1), (3,2), (4,3), gives an optimal solution with $a_{11}=h_b-J, a_{13}=\tfrac{J}{2}+h_b, a_{14}=\tfrac{J}{2}$, and the rest zero.  Here bouncing cannot be avoided, though in practice one could further shift the Hamiltonian which we have not done here for the purpose of illustration.  We note that we have just chosen these two limiting cases to illustrate the method.  In reality the simplex rules are applied for any numerical values of the couplings.

\subsection{Cycling}
\label{appsub:cycling}

We would finally like to comment on one possible issue that could arise in the general use of this algorithm, although we have not encountered the problem in our usage.  For convenience we have always assumed a $4 \times 4$ matrix of update moves, where some matrix elements could be zero.  We see that pivoting on these rows does not change the value of the objective function, and it could happen that a sequence of moves returns the tableau to a previous state.  This would result in an infinite loop where the simplex algorithm fails to terminate.  

This can be remedied relatively easily by increasing the size of the simplex tableau.  One regards the elements in the last column of the tableau as being row vectors, and replaces those initial elements with $W_1 \to (W_1,1,0,0,0)$, $W_2 \to (W_2,0,1,0,0)$, $W_3 \to (W_3,0,0,1,0)$, $W_4 \to (W_4,0,0,0,1)$, $-\sum_i W_i \to (-\sum_i W_i,0,0,0,0)$.  We then compare vectors by their lexicographical order, meaning $\mathbf{W}_i < \mathbf{W}_j$ if the first differing element between the two is smaller for $\mathbf{W}_i$.  For example (0,1,1,3,4) $<$ (0,1,2,2,4).

Now the only change to the simplex algorithm is that when we select the pivot row, we must choose the pivot $p_{ij}>0$ with the lexicographically smallest $\tilde{\mathbf{W}}_i/p_{ij}$.  It can be shown that pivoting according to the rules of the simplex algorithm always increases the lexicographical value of the objective function.  Since the value always increases, this process must terminate with an optimal solution, else the problem is unbounded, which is never the case here.

It also seems likely, and certainly more simple, to avoid cycling by choosing possible pivot columns randomly as opposed to sequentially.  Again, we have not needed to take such precautions as this issue has not come up in the various contexts where we have used this algorithm.
\end{document}